\documentclass{IEEEtran}
\usepackage{cite}
\usepackage{graphics,graphicx,amssymb,amsmath,amsfonts,verbatim}
\usepackage{textcomp,nicefrac}
\usepackage{epsfig}
\usepackage{subfigure}
\usepackage[subnum]{cases}
\usepackage{color}
\usepackage{lineno}
\usepackage{epstopdf}
\usepackage{bm}

\newcommand{\Figure}[1]{\textit{Figure}~#1}

\newcommand{\expectation}[2]{\ensuremath{\mathbb{E}_{#1} \left[ #2 \right]}}

\newcommand{\KLD}[2]{\ensuremath{ \,\mathrm{KL}\left( #1\lVert#2 \right) } }

\def\BibTeX{{\rm B\kern-.05em{\sc i\kern-.025em b}\kern-.08em T\kern-.1667em\lower.7ex\hbox{E}\kern-.125emX}}
\begin{document}

\title{Diagnostic data integration using deep neural networks for real-time plasma analysis}

\author{\IEEEauthorblockN{A.~Rigoni Garola, R.~Cavazzana, M.~Gobbin, R.S.~Delogu, G.~Manduchi, C.~Taliercio, A.~Luchetta}
\thanks{A. Rigoni Garola (andrea.rigoni@igi.cnr.it) and all authors are with \textit{Consorzio RFX (CNR, ENEA, INFN, Universit\'a Degli Studi di Padova, Acciaierie Venete SpA)}, Corso Stati Uniti 4, Padova 35127, Italy. \newline Submitted: October 2020 for the \textit{22nd IEEE Real Time Conference}}}

\maketitle

\begin{abstract}

Recent advances in acquisition equipment are providing experiments with growing amounts of precise yet affordable sensors. At the same time, an improved computational power, coming from new hardware resources (GPU, FPGA, ACAP), has been made available at relatively low costs. This led us to explore the possibility of completely renewing the chain of acquisition for a fusion experiment, where many high-rate sources of data, coming from different diagnostics, can be combined in a wide framework of algorithms. 
If on one hand adding new data sources with different diagnostics enriches our knowledge about physical aspects, on the other hand the dimensions of the overall model grow, making relations among variables more and more opaque. A new approach for integrating such heterogeneous diagnostics, based on the composition of deep \textit{variational autoencoders}, could ease this problem, acting as a structural sparse regularizer. 
This has been applied to RFX-mod experiment data, integrating the soft X-ray linear images of plasma temperature with the magnetic state.

However, to ensure a real-time signal analysis, those algorithmic techniques must be adapted to run in well suited hardware. In particular, it is shown that, attempting a quantization of neurons transfer functions, such models can be adapted to run in an embedded programmable logic device. The resulting firmware, approximating the deep inference model to a set of simple operations, fits well with the simple logic units that are largely abundant in FPGAs. This is the key factor that permits the use of affordable hardware with complex deep neural topology and operates them in real-time.

\end{abstract}


\section{Introduction}\label{section:Intro}
\IEEEPARstart{M}{any} successful approaches in making systems "smart" are not based on complicated mathematical models, but rather on simple functions and large amounts of data.
This intelligence emerges from optimization and statistics, eventually improving the knowledge of a complex environment by means of an approximation. 
This is clearly the object of the deep neural networks (DNN) in both supervised learning, where the network is trained to behave like an external model, and in unsupervised learning, where the network infers the characteristics of the dataset by itself.
In addition, recurrent neural networks added time as an additional model variable, opening the door to this optimization framework for dynamic control~\cite{bensoussan2020machine}.
Looking at the recent promising results in this area, it is not unrealistic to think of a neural network for the control of a complex non-linear system like a fusion experiment.
~
In this context the \textit{RFX~consortium} hosts \textit{RFX-mod}, one of the most controllable experiments currently active in fusion research~\cite{SONATO2003161}\cite{doi:10.1063/1.4806765}.
A further upgrade to this experiment (\textit{RFX-mod2}) is ongoing, aiming to improve the responsiveness of the machine, together with the renewal of the detector system~\cite{Marrelli_2019}.
The present work describes a design showing how DNNs could be integrated in the overall chain for data acquisition and control (CODAS), and in particular how they could be used to merge together heterogeneous signals coming from different diagnostics.

Being artificial neural networks inspired by brain functioning, an intuitive example of how they could be used in a control system could be suggested by the analogy with how we learn to perform a relatively simple task, like for example to ride a bike. No one really applies a physical description of the system to do that, and learning to maintain the balance on the two wheels is not about simply looking a posteriori if we felt or we stayed on path, and learn an optimal sequence of movements, but it comes from shaping a safe equilibrium domain in an inner reduced state space.
This state merges together not only the unknown physical aspects of the bike, but also all the input signals from complex non linear sensors ( vision, equilibrium, touch ) and the feedback itself ( balance, steering handlebar, pushing on pedals ).
Two intuitive proofs come again from experience: for example, if we close our eyes for a while we will still be able to maintain the equilibrium, because we use other senses to continue the reconstruction of the inner state. Although if we try to switch the hands on the handlebar we will probably fall in a few meters, even if the system remains the same, because the network receives inverted inputs of the learnt feedback and the state goes immediately out of the control area.

Just like the bike example, the difficulty to control a plasma experiment could reside not much on the plasma itself, but in the access of the real state of the machine because of the complex non linear models that link all kinds of diagnostics. 
Currently a very small amount of sensors are exploited in \textit{RFX-mod} as inputs for the control system, the bulk of acquisition being left as mere diagnostics for the off-line analysis.
~
The purpose of this work is, in fact, not to demonstrate the feasibility of a complete machine learning control system for a fusion experiment yet. But it is proposed as a preparatory step that aims at finding an effective method to provide a representation of the system state enriched by diagnostics information that has never been integrated in real-time before.
A complex and heterogeneous experimental system could still benefit from this methodology in many ways: the possibility to integrate different kinds of signals, the robustness to react to missing or corrupted data in case a portion of the sensory system fails, and the flexibility to adapt to changes on both the parameters or the environment variables of an experiment where all the physical aspects are not tractable in real-time, or even they are not completely known.

In the following: \textit{section}~\ref{section:2} introduces the neural network structure used to perform such system integration. In \textit{section}~\ref{section:3} a simple test case is presented based on RFX-mod measures of the internal profile of the plasma electron temperature ($T_e$) provided by the Soft X-ray (SXR) diagnostic. This signal will be then integrated with external measures composed by some plasma parameters and the magnetic state. 
Another network, mapping these two inputs, revels that this approach has been able to transfer the common hidden factors that link the plasma temperature to the magnetic configuration in \textit{section}~\ref{section:4}--{section:5}
This characterizes a more general state of the experiment that can be used to further improve the former signal, as explained in \textit{section}~\ref{section:6}. A final proof of concept hardware implementation is reported in \textit{section}~\ref{section:7} showing the feasibility of time constraints for a possible control scenario.

%
%
%
%
%
%
%
%
%
\section{Heterogeneous Diagnostics Integration (HDI) using neural networks} \label{section:2}
\begin{figure}[t]
\centering
\includegraphics[width=0.28\textwidth]{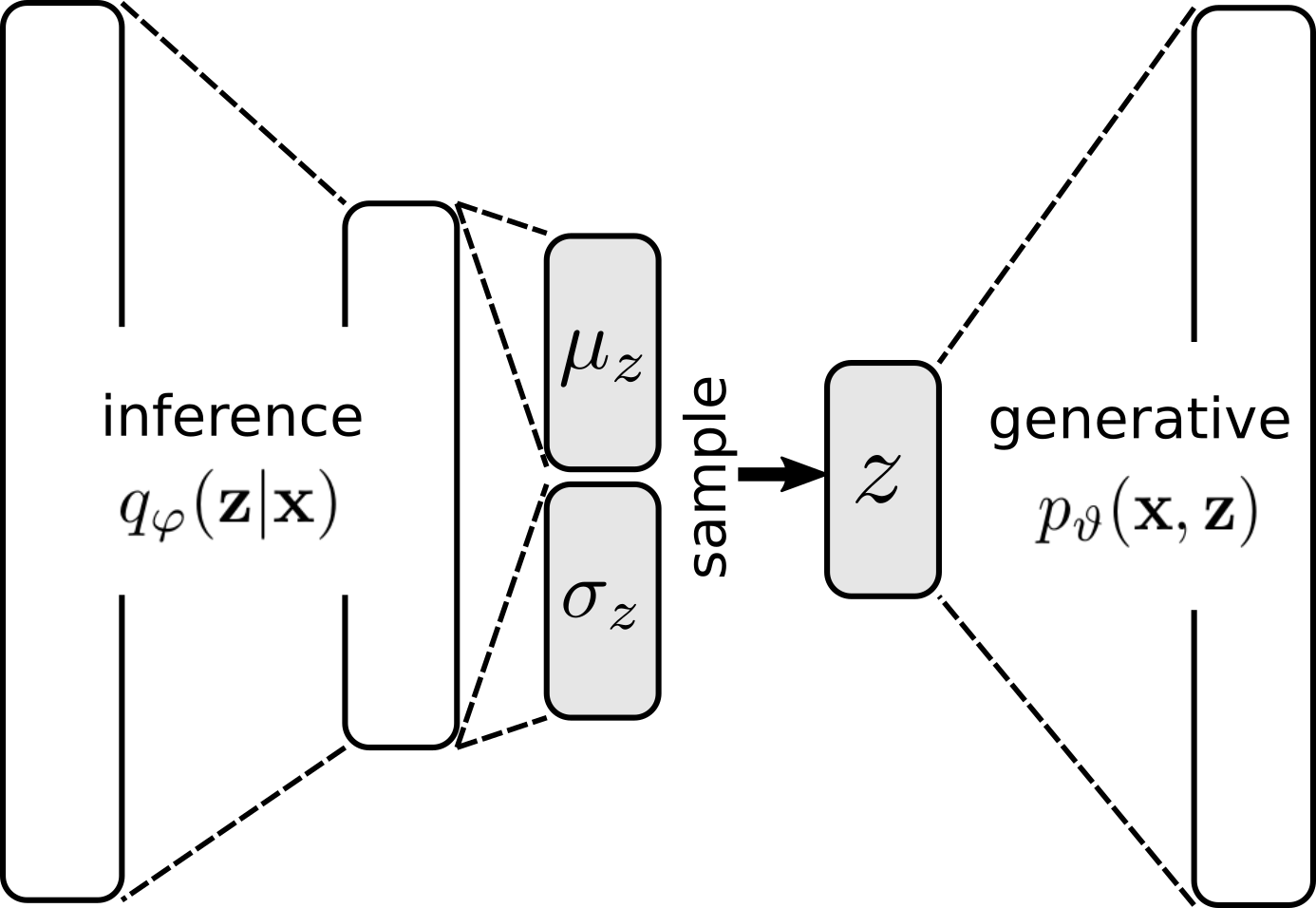}
\caption{General schema of a variational autoencoder.}
\label{fig: vae}
\end{figure}

~
In almost every branch of science the process of understanding a phenomenon goes thought the creation of a model, unveiling the world from hypotheses and testing these through observations.
But instead of simply building a predictor that fits observations into given physical hypotheses ( inference modelling ), we propose to exploit neural networks as a black box to simulate how the data has been generated in the real world ( generative modelling ). 
For this purpose we will rely on the network topology called \textit{variational autoencoder} (VAE)~\cite{DBLP:journals/corr/KingmaW13,DBLP:conf/iclr/2014}. 
~
The simple autoencoder is a network structure characterized by input and output layers of the same size, and falls in the category of unsupervised learning algorithms as it tries to reconstruct an exact match of the inputs once passed through the network non-linear transfer function.
An autoencoder generally consists of two parts: an encoder which transforms the input to a hidden code, and a decoder which reconstructs back the input from it. The outcome is a reduced dimension on at least one of the hidden layers, that is the output of the encoder. This induces the network to extract the main representative characteristics of the data~\cite{Goodfellow-et-al-2016}~\cite{murphy2013machine}.
%
%
Variational autoencoders add a probabilistic meaning to the internal states treating them as pure stochastic entities called \textit{latent variables}.
This turns the model to be a representation of the joint probability distributions and high-order dependencies between the input and the latent space. If we assume the variable $\textbf{x}$ to be a random variable $p^*(\textbf{x})$ from the unknown process, the model becomes an approximation in terms of a set of parameters $\pmb{\theta}$:
\begin{equation}
p_\theta(\textbf{x}) \simeq p^*(\textbf{x}).
\end{equation}
Latent variables $\textbf{z}$ are the unobserved part of this model, linked with each output data by the \textit{marginal likelihood}, also called the \textit{evidence}:
\begin{equation}
p_\theta(\textbf{x}) = \int{p_\theta(\textbf{x},\textbf{z}) d\textbf{z}}.
\end{equation}
We are looking for an inference model that creates a distribution $q_\phi(\textbf{z}|\textbf{x})$ that maximizes the log-likelihood of the model reconstructed data $\log p_\theta(\textbf{x})$.
It can be shown~\cite{DBLP:journals/corr/KingmaW13} that this is equal to maximize a term named \textit{evidence lower bound} (ELBO):
\begin{equation} \label{eq:elbo}
\begin{split}
    \mathcal{L}_{\theta,\phi}(x) & = 
    \log p_\theta(\textbf{x}) - \KLD{q_\phi(\textbf{z}|\textbf{x})}{p_\theta(\textbf{z}|\textbf{x})} \\
    & =  \expectation{ q_\phi(\textbf{z}|\textbf{x})}{ \log p_\theta(\textbf{x},\textbf{z}) - \log q_\phi(\textbf{z}|\textbf{x}) } 
\end{split}
\end{equation}
It can be seen as a lower bound because it is defined as the difference between the log-likelihood and the positive definite term described by the Kullback-Libeler (KL) divergence between the two posterior distributions over $\textbf{z}$: the inference posterior $q_\phi(\textbf{z}|\textbf{x})$, and the true posterior $p_\theta(\textbf{z}|\textbf{x})$.
On the second equation the first term inside the expectation is the distribution output of the generative model, while the second term is from the inference model. If a DNN network is used to implement such models $\theta$ and $\phi$ will represent the trainable parameters of respectively the generative and the inference networks, as shown in \textit{Figure}~\ref{fig: vae}.
The learning proceeds through a minimization of a loss that is then the opposite of this function:
\begin{equation}
    \theta^*,\phi^* = \text{arg}\min_{\theta,\phi} \left( - \mathcal{L}_{\theta,\phi}(x) \right)
\end{equation}
However the gradient on $\phi$ is computationally intractable, due to the expectation operation.
For this reason a reparametrization of $\textbf{z}$ is applied by a further function: $\textbf{z} = g(\textbf{x},\varepsilon,\phi)$, where $\varepsilon$ is a random independent value generated from the output of the inference network.
\begin{equation}
    \nabla_\phi \expectation{q_\phi(\textbf{z}|\textbf{x})}{f(\textbf{z})} = 
                   \expectation{q_\phi(\textbf{z}|\textbf{x})}{ \nabla_\phi f(\textbf{z})} \simeq
                   \nabla_\phi f(\textbf{z})
\end{equation}
A common choice is to sample $\textbf{z}$ from a set of factorized Gaussian distributions:
\begin{equation}
\varepsilon \sim \mathcal{N}\left(\textbf{z};\pmb{\mu},\text{diag}(\pmb{\sigma}^2)\right)
\end{equation}
A sketch of this process is reported in \textit{Figure}~\ref{fig: vae} where both terms of ELBO are shown. The reparametrization parameters $\pmb{\mu},\pmb{\sigma}$ are inferred from the input data, and used to sample a new $\textbf{z}$ value. Then this particular value is eventually generating the output.
In this way an autoencoder can be defined as a neural network whose primary purpose is to learn a lower dimensional underlying manifold of the features in the dataset, i.e. the ``\textit{latent space}''. 
But, unlike other dimension reduction methods, the autoencoder is not based on a single property of the data like distance (k-means, t-SNE)~\cite{Kanungo00anefficient,vanDerMaaten2008}, or on the data topology (PCA, LLE)~\cite{doi:10.1080/14786440109462720,Roweis2000}.
The characteristic of VAE is that, on one end, training the encoder we optimize the model parameters to fit the given dataset, but, on the other end, we also emphasize the causal factors that transforms these latent variables back into the generated samples~\cite{Kalainathan2019GenerativeNN}.
In addition the \textit{equation}~(\ref{eq:elbo}) can be also expanded as:
\begin{equation}
    \mathcal{L}_{\theta,\phi}(x) = 
    \expectation{ q_\phi(\textbf{z}|\textbf{x})}{ \log p_\theta(\textbf{x}|\textbf{z})} - 
    \beta \KLD{q_\phi(\textbf{z}|\textbf{x})}{p_\theta(\textbf{z})}
\end{equation}
The first term is exactly the likelihood of data conditioned by the encoded state, and it is trained by the standard backpropagation error, i.e. mean squared error (MSE) between input and output. The second term is instead a Kullback–Leibler divergence between the inference distribution and the prior. This term (analytically computed) acts as a ``\textit{regularizer}'' that makes the system generalize better on unseen states, and promotes the so-called \textit{latent variables disentanglement}~\cite{Higgins2017betaVAELB}. 
We also applied a tuning factor $\beta$, to keep under control the regularization effect that tends to be over-estimated on real data.

The neural networks approach represents a very appealing method for creating a precise approximation of these non-linear models using simple elements, discovering hidden relations among many different inputs. 
However, since this process passes through the optimization of many parameters, a common issue is that obtaining a good fit of the model from an increase of the input dimensionality requires an exponential growth of the training data. This is a well known aspect of numerical computation usually referred to as ``\textit{the curse of dimensionality}'' ( term coined by R.~Bellman in ``\textit{Dynamic Programming}'' 1957 ), and it has been challenging the control of complex systems for quite a long time.
~
For this reason, instead of feeding all possible acquired measures in a single big network, we propose a set of many smaller VAEs that are specifically tailored to fit data from each of the involved diagnostics. The final integration of different sensors happens with the composition of the encoded signals in a further VAE. This mixes these encoded latent variables into a general state of the machine, highlighting even more the main causal factors of the overall system.

From a general perspective this corresponds to create a structural sparse approximation of the whole experiment, where the relations between single inputs of different diagnostics is discarded in favor of the high level composition of their encoded latent representations.
~
This provides the correct dimensionality for successful training and, at the same time, the network itself can be designed to fit within the diagnostic device on the edge of the acquisition system, benefiting in this way of the high data bandwidth from the sensor.

\begin{figure}[ht]
\centering
\includegraphics[width=0.48\textwidth]{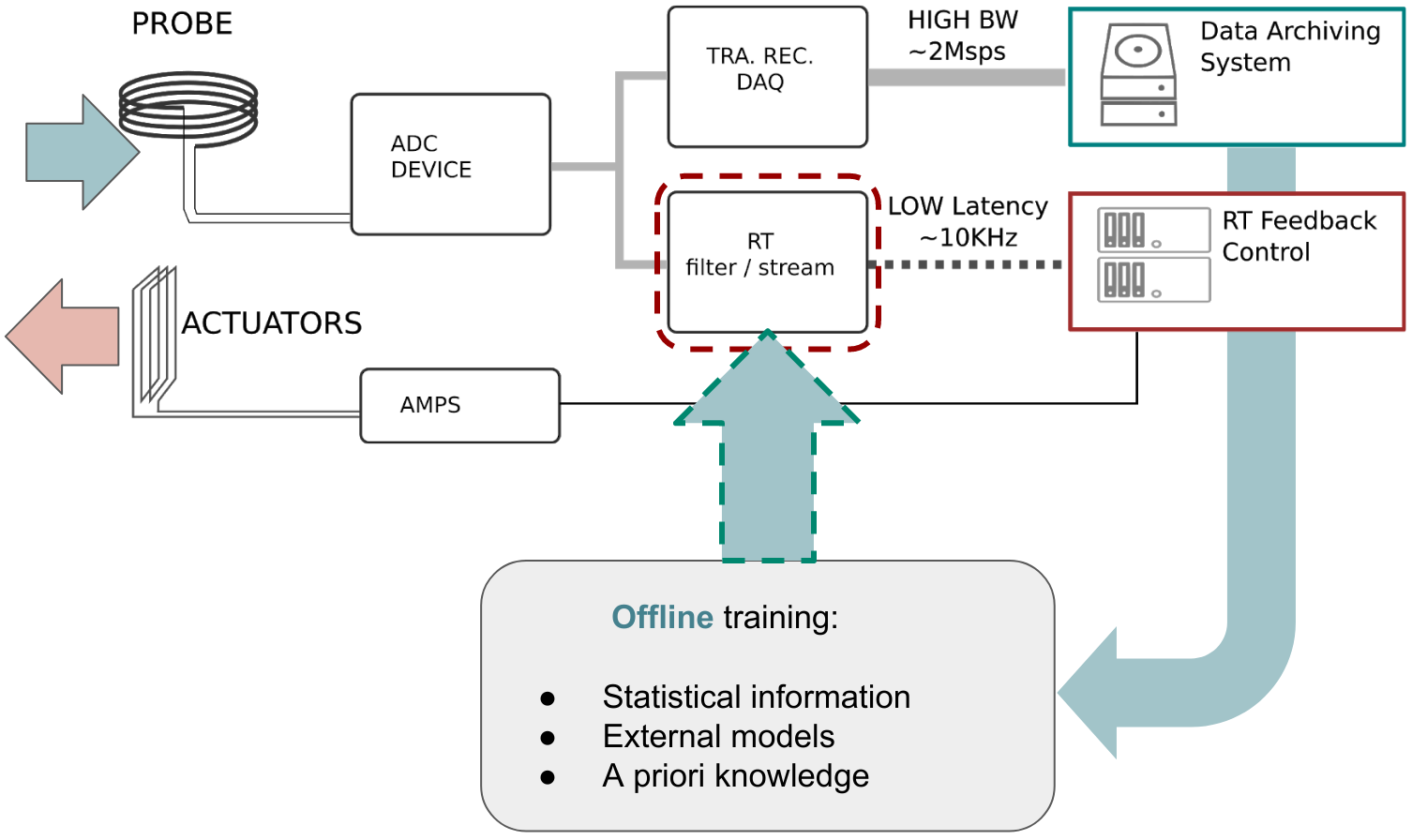}
\caption{Logic design of the flexible ADC structure where the stored dataset is used to train an encoder offline. The trained network can then be deployed as a special FPGA filter that sends codes through a low latency network to a higher level encoder, or to the main control system. }
\label{fig:1 logic}
\end{figure}
The typical internal organization of a generic experiment CODAS is sketched in \textit{Figure}~(\ref{fig:1 logic}). The input acquisition from a sensor probe is preconditioned and digitized by the acquisition device (DAQ) that is further divided in two main components: a transient recorder acquires high data rate signals that are buffered in local memory and sent to the main data storage for offline data analysis, at the same time a sub-sampled signal with a low latency channel is sent to the central control unit as input of the feedback model.
We propose to improve or replace this latter sub-sample operation with the output of an encoder network, trained from experiment offline dataset and deployed in FPGA to optimize latency. This can be used: to add much more information to the low-latency channel, to embed acquired statistical information from the experiment, and to inject a priori knowledge on data, directly to the real-time control.
~
We will also show that this can be composed with the encoded output of other diagnostics to create a better representation of the originating signal from the device, effectively recovering from noise and even missing data. 
%
%
%
%
%
%
%
%
\section{A Test case based on plasma soft X-Ray temperature profiles} \label{section:3}
In order to demonstrate the feasibility of the proposed approach in a real plasma environment,
we considered the electron temperature ($T_e$) profiles that are produced by the Soft X-Ray (SXR) diagnostics in \textit{RFX-mod}~\cite{Franz_2001}~\cite{franz2013experimental}. 
The SXR diagnostic named \textit{DSX3} is essentially a X-ray camera installed in one of the equatorial ports of the experiment and acquires the integrated SXR emissivity of plasma along 18 lines of sight that cross the toroidal section of the chamber. This brightness is hence related to the electron temperature of the plasma by means of bremsstrahlung radiation.
~
We know that the \textit{RFX-mod} experiment \textit{reversed field pinch} configuration (RFP) is keen to produce a large amount of magnetic instabilities due to the low profile of the safety factor. These current driven instabilities can grow in the core of the plasma producing the so-called ``\textit{magnetic islands}''~\cite{Cappello_1996}.
At the islands toroidal position the plasma emits more, and localized regions are identified by SXR producing different temperature profiles always shaped as convex curves: they can be either almost symmetric, if many magnetic instabilities overlap each other (\textit{multiple helicity} MH), or intrinsically asymmetric, if one perturbation dominates (\textit{quasi single helicity} QSH) showing a protrusion that corresponds to the phase of the helix at the camera toroidal position.

The particular choice of this diagnostic had been mainly motivated by the fact that it provides a realistic representation of the plasma configuration, being at the same time not too complex to be handled by simple networks. It can be also seen as an instant property (almost ergodic during the pulse), so there is no need to implement a dynamical system to match this measure with the magnetic state, and acquired with a good temporal resolution (about $500~\mu s$ in \textit{DSX3}).
But, if SXR gives a good local information of the internal state of plasma, giving a chance to look beyond the simple external magnetic configuration, it is also severely affected by noise that can even overflow the camera sensitivity at some pixels, in particular because of plasma wall interaction.

We studied the possibility to recover this signal, first using the statistical information coming from the same signal previously acquired, and then integrating an external state of the system as seen from magnetic sensors.
The idea is to see if a neural network is able to learn how the plasma internal temperature links with the magnetic state, even if a complete model of this matching would require a lot of complex physical assumptions and a very high computational power~\cite{Martines_2011}.

The study has been divided in three steps that are described in the following sections.
In \textit{section}~\ref{section:4} a variational autoencoder has been applied to improve the output of \textit{DSX3}, recovering from noise and missing input data from the sole experimental statistics. The \textit{section}~\ref{section:5} will show how the state space generated by this autoencoder can also be used to map the soft X-ray data to external diagnostics, i.e. the magnetic configuration measured by in vessel coils and some parameters like the plasma current, theta, and the reversal factor. All the data from both these apparently independent measures are then concurring to build a single state space that further improves the reconstruction of the temperature profiles in \textit{section}~\ref{section:6}. 

%
%
%
%
%
%
%
%
\section{STEP1 - Soft X-Ray signal recovery} \label{section:4}
~
The dataset we used comes from an extensive experimental campaign of RFX-mod with many configurations and plasma currents that reach the $1.5\times10^6\text{A}$~\cite{Gobbin_QSH}.
Each point of the input profile is composed of two values: the impact parameter on x-axis that is the perpendicular distance between the line of sight and the center of the chamber section, and the measured temperature on y-axis. The points positions along x-axis is not fixed on the dataset because they depend on the position of the camera that can be adjusted.
Within this temperature dataset the majority of samples are actually missing in some of the acquired positions. This means that value for the corresponding line of sight has not been recorded because too noisy in that moment and have been assigned to NaN (Not a Number IEEE-754).
~
Moreover, the distribution of these missing points appears to be strongly not-uniform, because some of the acquired positions are so noisy that we do not have enough statistics to train the network effectively. For this reason we decided to apply a shrink of the dataset from 18 positions to 15 using a k-means approach. We defined $k=15$ centroids and the k-means was fitted on the basis of the x-values (i.e. the position of the impact parameter). This is actually loses some of the available information on the few input data that present more than k elements; nevertheless it allows to have a distributed set of positions to correctly feed all neurons of the network with almost the same statistic.
Then all impact parameters and the temperatures (i.e. $x$ and $y$ values) have been fed in a VAE build on simple feed forward layers leaving the network to associate them in the profile $(x,y)$ points.

That said, we are still facing a sparse input feature set where many values are NaN and does not comply with internal operations that implement neural networks. The common approach is to fill the empty gaps with a pre-computed values before feeding the signal in, this is commonly referred to as \textit{data imputation} preconditioning. Many algorithms exist, and a general accepted approach is a regression of some kind in the region adjacent to the missing data. This seemed a well documented and easy way to go at first glance; however in this dataset we are facing many adjacent missing points and a "simple" regression (polynomial or ridge) didn't produce a good fit for all the samples. 

To cope with this problem we adopted an approach derived from the dropout technique commonly adopted to improve generalization in Neural Networks. 
\begin{figure}[ht]
\centering
\includegraphics[width=0.48\textwidth]{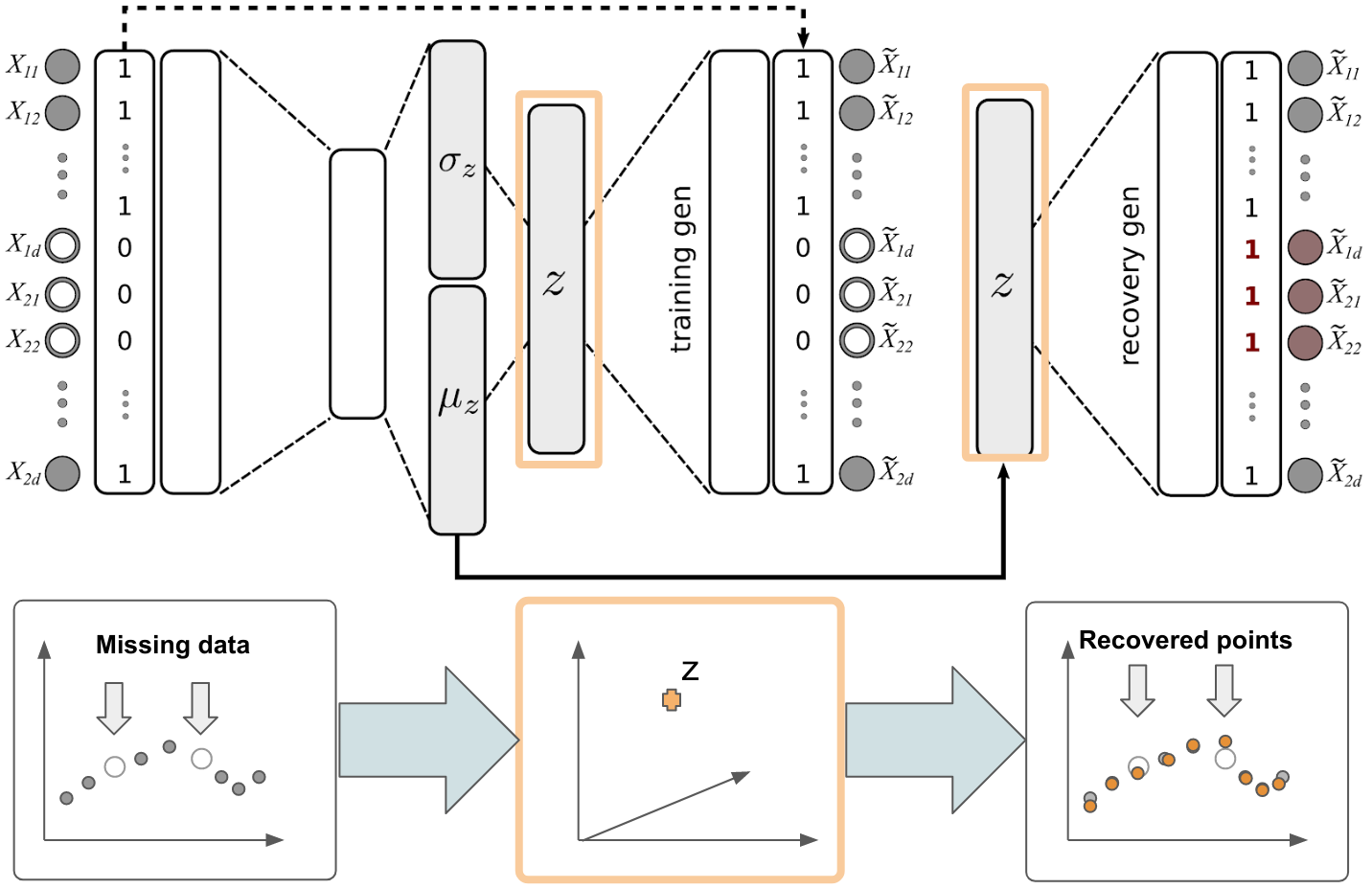}
\caption{Variational autoencoder structure adapted for the recovery of missing points. A general structure of VAE is inserted between two mirrored layers, the \textit{NaN-mask}, that prevent the training on the missing paths. On the lower part of the image the recovery process is done creating a value for $z$ in the latent manifold, and recovering the signal back with the last layer set to 1.}
\label{fig:2 recover}
\end{figure}
The solution is sketched in~\textit{Figure}~\ref{fig:2 recover} where two mirrored layers on the head and on the tail of the overall network have been added to the legacy VAE schema. In particular the layer, called \textbf{NaN-Mask}, transforms all NaN to zero floats on the inference network, meanwhile the same mirrored transformation is applied on the network output erasing the generated reconstructed point. The weights of the added layers are not changed by the learning procedure. This is equivalent to randomly switching off a subset of inputs during learning. 
It turns out that training succeeds even with the missing data, implicitly using the other available information, and the missing inputs, instead of being just only an issue, add in this way a regularization effect on the training (input \textit{drop-out}) that slightly helps to prevent overfitting~\cite{dropout}.

During the actual reconstruction the NaN-Mask leaves unactivated all neurons facing the missing data, and the corresponding input is treated as noise by the encoder that still tries to create a point in the latent space. This point is then reconstructed back by the generative network with all active outputs this time, generating a statistical best guess for the NaN values.
Several topologies have been tested and we got a good reconstruction using a symmetrical set of four \textit{ReLU} layers with 300 nodes per layer and a latent space of 6 dimensions ($\text{VAE}_6$ hereafter).
%
%
%
%
%
%
%
%
%
%
\section{STEP2 - Magnetic configuration to electron temperature mapping} \label{section:5}

\begin{figure}[ht]
\centering
\includegraphics[width=0.48\textwidth]{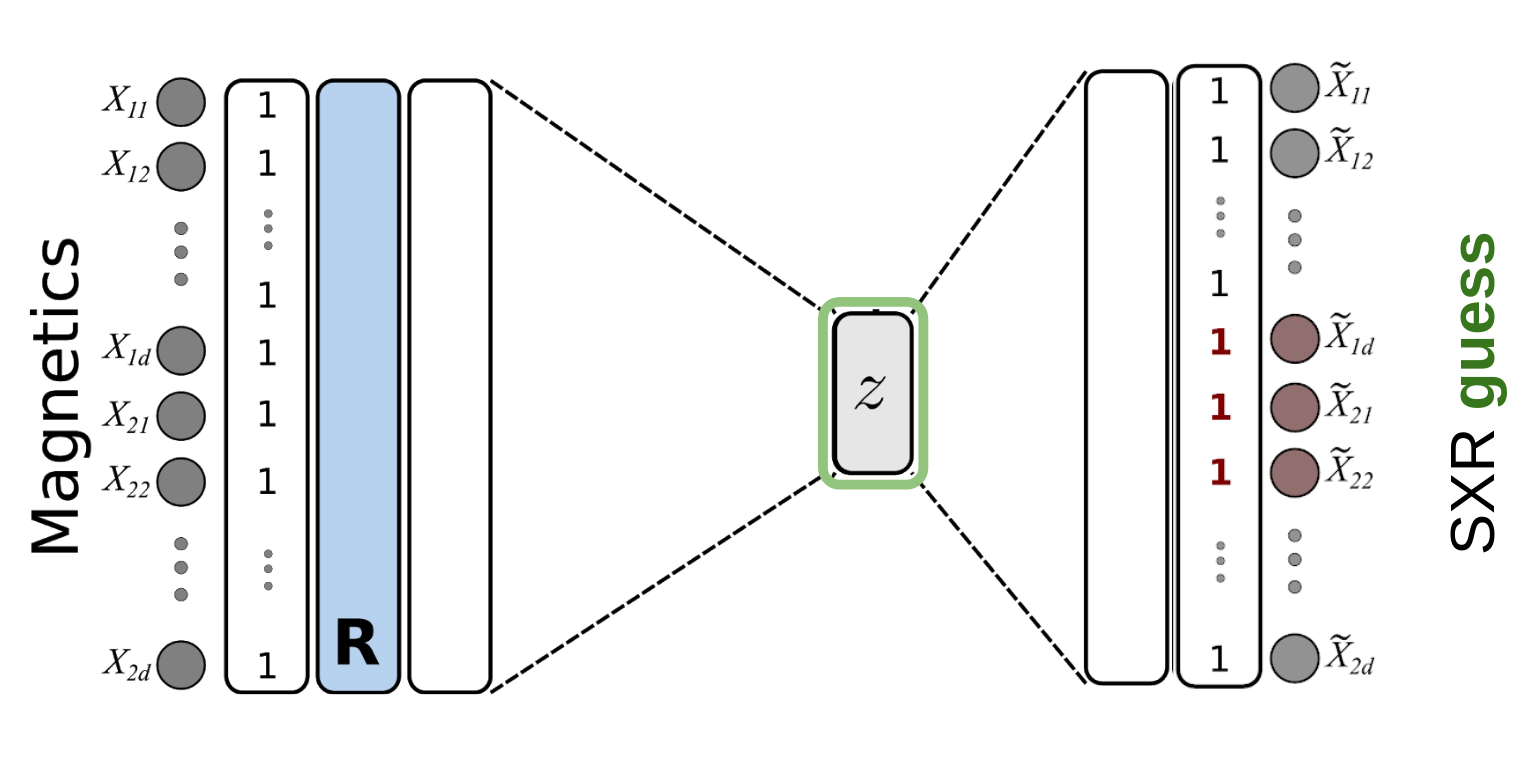}
\caption{Topology of the network that maps external information to the latent temperature profile representation. The complete mapping is provided by a inference network that guess the state of the temperature from the magnetic status and by the autoencoder generative network that decode the profile guess.}
\label{fig:3 map}
\end{figure}

In the introductory section we argued that the information redundancy brought by different diagnostics on the same system internal state would increment the robustness of the implemented control over a possible failure of a sensor or to recover added noise.
~
As a proof of this idea we will show that, even if a very complex physical model links the plasma temperature to the magnetic configuration, a relation can be effectively approximated by DNN.
The principle would be to let the network create these relations by itself, instead of guessing a possible affinity with one quantity or the other.
The schema will not be much more complex as it has been so far: the autoencoder shapes a manifold in the latent space from where we take values to train the magnetic input dataset fed into a standard supervised network. 
A sketch of such implementation design is reported in~\textit{Figure}~\ref{fig:3 map}.
The external 24 input values here are: the absolute value and phase of the magnetic modes characterized in the spatial Fourier transform by poloidal number $m=1$ and toroidal numbers $n=7 \to 16$, the plasma current ($I_p$), the ratio of the dominant mode ($\text{NS}$), the pinch parameter ($\Theta$), and the reversal parameter ($F$).
An inference network is then trained with this input to match the latent state corresponding to the temperature profile acquired by \textit{DSX3} at the same time. The mapping is then completed by the former autoencoder generative network trained at the first step. 
~
The network topology that has been exploited had four \textit{ReLU} feed forward layers (two with 880 nodes and two with 440 nodes) and it has been trained with a dropout of $0.05$. There is no missing data among these input parameters so all the \textit{NaN-Mask} weights are set to $1$, but a further ``R'' layer is present. This is a special layer where the weight matrix has been constrained to be kept diagonal, so after the completion of training the values of these weights represent the mutual information of that particular input and the reconstructed state. It gives a chance of looking at which are the most relevant parameters that link one set to another, or the magnetic modes to the temperature in this case. A more detailed description is left for a future publication, although this is a fully trainable layer so it must be mentioned here because it affects the reconstruction. 

\begin{figure}[ht]
\centering
\includegraphics[width=0.48\textwidth]{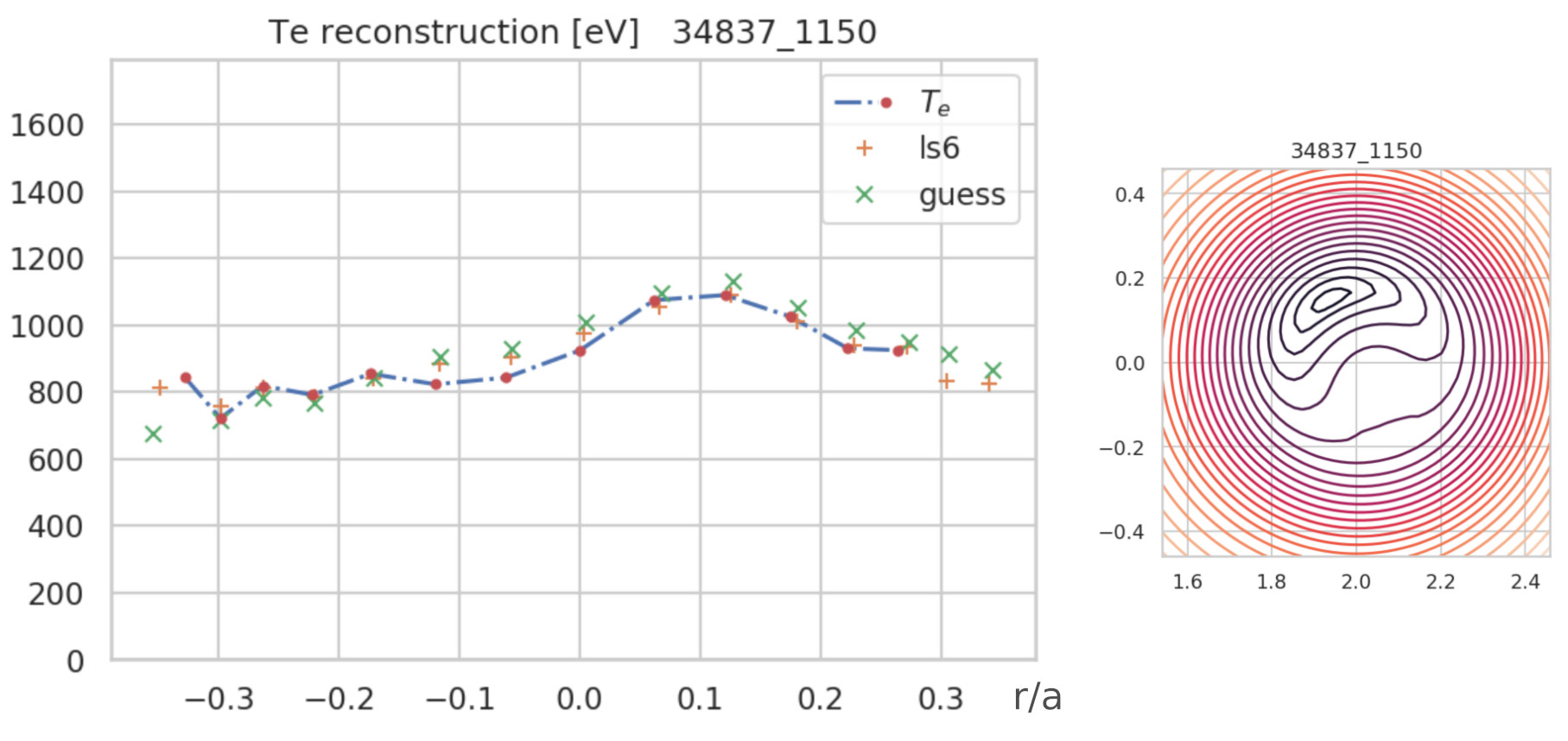}
\caption{Temperature profile reconstruction example (left): the real $T_e$ input ($\cdot$), $\text{VAE}_6$ reconstruction (+), and the guess of external magnetic information ($\times$). The SHeq reconstructed magnetic profile (right).}
\label{fig:5 map example}
\end{figure}

An example of reconstruction of these signals is shown in \textit{Figure}~\ref{fig:5 map example};
the real $T_e$ profile is represented by the blue dashed line that is interrupted where its values are missing, all available points are instead represented as red dots ($\cdot$). The simple $\text{VAE}_6$ reconstruction that shows the missing data recovery explained in \textit{Section}~\ref{section:4} is marked with an orange cross (+). The external reconstructed profile that comes form a complete external guess of the state is instead represented by green diagonal crosses ($\times$). This shows how the network was able to reproduce the temperature characteristic of the plasma using a complete external information. And it can be done in real-time, whereas a numerical computation based on a physical model (that is not yet completely descriptive) would require hours of CPU time. The 2D plot on the right represents the magnetic configuration of the corresponding section of the plasma that comes from numerical simulation \textit{SHeq}~\cite{Martines_2011}.

%
\section{STEP3 - Improving signal recovery adding magnetic state external information} \label{section:6}
In \textit{Section}~\ref{section:4} and \ref{section:5} data that has been accounted is either the temperature profile or the plasma parameters, each of them being considered as a single instance. The neural network approach shows that it is possible to learn how to transfer the common hidden factors that drive experimental outcomes from one diagnostic to another, and in particular a latent space embedding helps to match such information.
However, aiming at finding a comprehensive representation of the system, a more favourable approach would be to integrate those information together in a unique latent "state" space.

\begin{figure}[ht]
\centering
\includegraphics[width=0.48\textwidth]{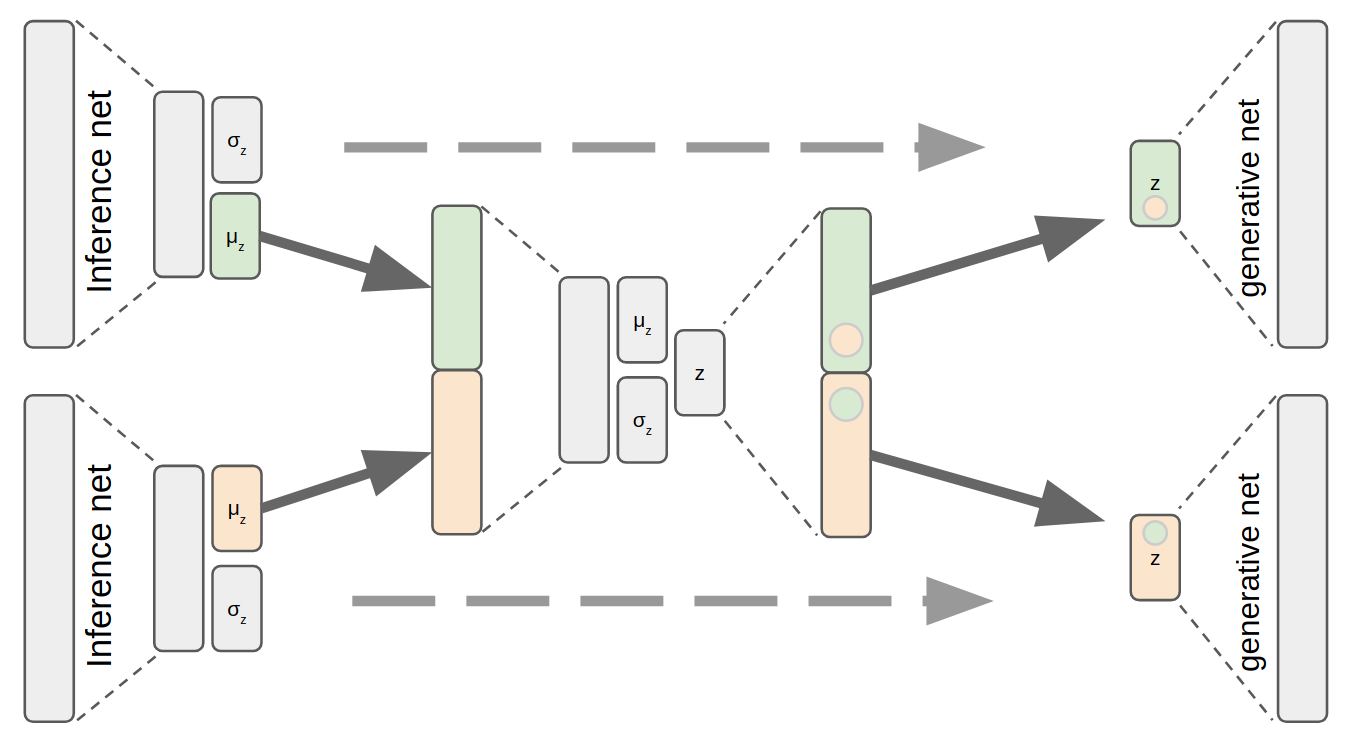}
\caption{A sketch of the latent variables composing structure. After all diagnostic encoder have been trained, the output of their $\pmb{\mu}$ component is fed into a nested internal autoencoder that create a more general state from the composition.}
\label{fig:4 composing}
\end{figure}

The actual idea that follows the intuition of a hierarchical subdivision of the complete data inputs in sub domains, each represented by means of an embedded manifold, has been realized as a chain of nested autoencoders. This is actually as simple as a new class method that created an internal structure from a input list of further autoencoders; this has been named \textit{latent spaces composition}, and a schema of the built topology is proposed in~\textit{Figure}~\ref{fig:4 composing}.
The sketch shows an example of the composition of two sets of variables, in this case the \textit{DSX3} output and the magnetic parameters.  The composition proceeds in two phases: we first separately train each of the single diagnostic external autoencoder as if they were single instances. Then the training of these networks is stopped and we train the nested internal autoencoder with the reconstructed codes as input. 
~
Moreover, the loss function associated with each of the external autoencoder can output values of different magnitude, but at the second stage the codes are more similar to each other thanks to the regularization effect of VAE. If this would not be enough the composing autoencoder we developed can further adjust the loss of each separate code with a specific weight parameter.
\begin{figure}[ht]
\centering
\includegraphics[width=0.48\textwidth]{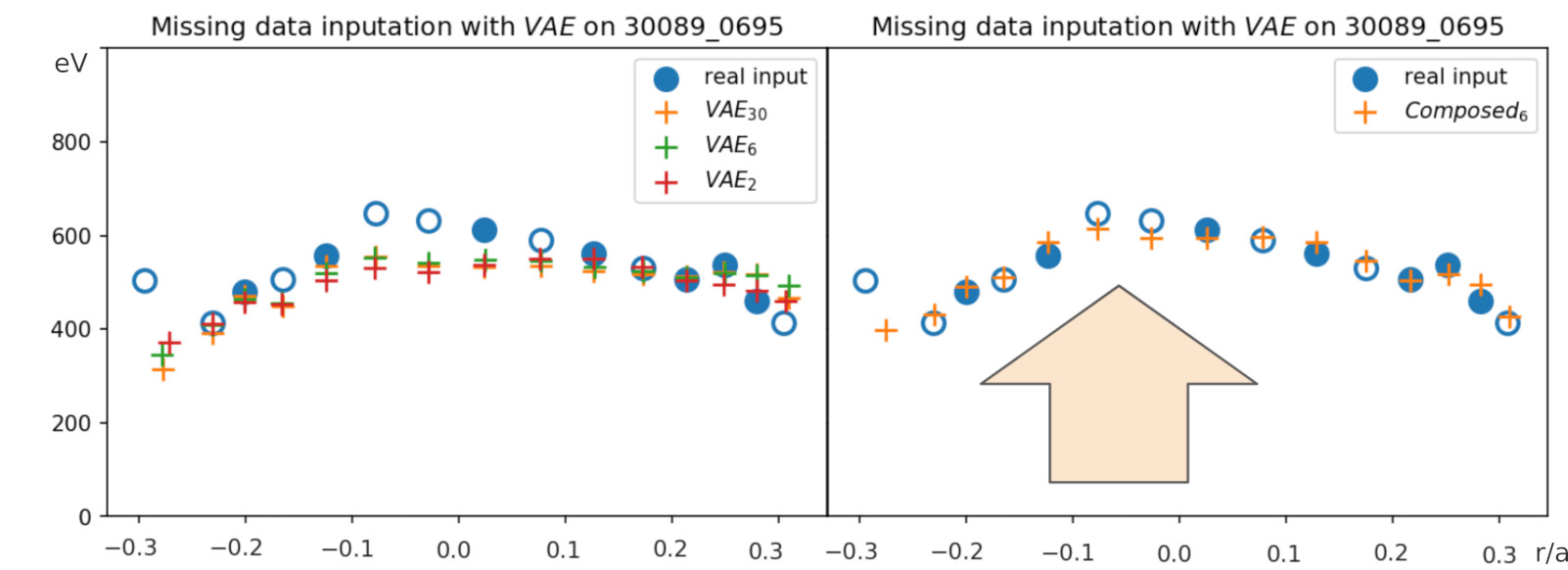}
\caption{An example of $T_e$ reconstruction where some points have been removed: the simple autoencoder fails to identify the QSH with any choice of latent dimension (left), the magnetic information that comes from the composite state remove this uncertainty and recover the actual shape of the profile (right).}
\label{fig:6 composing example}
\end{figure}
An example of this effect is proposed in \textit{Figure}~\ref{fig:6 composing example} where we deliberately removed some points on a temperature plot, testing the reconstruction of the temperature generative network: from the encoded values that come from the sole SXR diagnostic (left plot), and from the code that comes from the composition with the magnetics (right plot). The removed points create a special unfortunate case where the convexity of the curve is not clearly identified by the encoder because all of the points in the bulge that reveals this particular QSH are missing. In the first plot the encoder size has been set with different sizes (i.e. $2,6,30$) showing that the reconstruction error is not a matter of fitting data, in this case, but comes from a lack of information on the actual state of the system. On the contrary, in the second plot it is clearly visible that the magnetic data aids the reconstruction of the correct state of the temperature that now correctly identify the proposed QSH profile.

A quantitative result of the reconstruction error can be simply reported as the value of the loss function in the validation step, that is the mean squared error (MSE) of the input data and the reconstructed value of unknown inputs. 
The MSE of $\text{VAE}_6$ averaged on the validation dataset of 6000 samples reports an error of $0.6\times10^{-4}$ on normalized dataset, that corresponds to an average error of $17~\text{eV}$ per reconstructed point.
\begin{figure}[ht]
\centering
\includegraphics[width=0.48\textwidth]{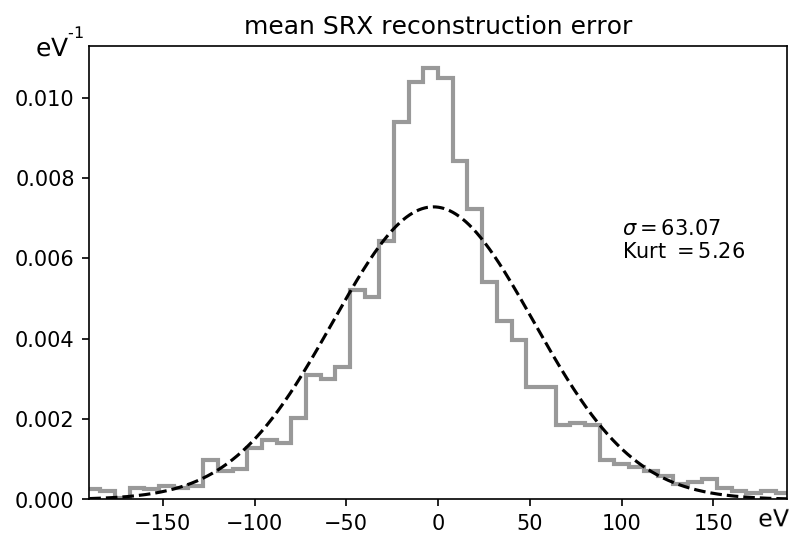}
\caption{Recovery of missing point error distribution where the missing value has been simulated from complete SXR profiles deliberately removing one randomly chosen point.}
\label{fig: err}
\end{figure}
It represents the result of both the recovery of missing point and the denoise effect. To see only the reconstruction contribution a missing data simulation has been set, removing one or many points randomly from complete profiles. The sole recovery distance from the real hidden point has been accounted in this case, and the error distribution is reported in \Figure\ref{fig: err} showing a zero mean leptokurtic shape with $\sigma=63~\text{eV}$.

%
\section{A proof-of-concept implementation: Low precision FPGA deployment in Zynq} \label{section:7}
If all these models were built by complex differential equations all the computation would need to be hardly optimized to fit the control timing constraints, and we would have to cope with numerical frameworks that are generally not keen to provide a real-time output.
But here comes the beauty of the approximation brought by DNN. A complex non linear model can be indeed fit into a set of simple operations. In practice, if we apply the simple ReLU activation, all reduce to a composition of linear functions.
~
This would be already a tremendous optimization, but we ask for more: we need the system to perform within the typical control time that is few $\mu s$, and we want the encoder to fit in the detector device to have the a low latency transmission of the code, instead of a decimated sample, from the diagnostics to the control system. This optimizes also the amount of exchanged information in the low latency network from the detector to the controller. With this goal in mind the best solution seems to deploy these trained networks in an embedded flexible FPGA unit, capable of being adjusted over time and reasonably cheap.
~
A good candidate architecture has been found in the embedded boards equipped with SoC units that share an ARM processing unit with a FPGA logic: the Xilinx Zynq 7k for instance. These boards match the optimal balance of flexibility, performance and costs required by our experiment and we are already planning to integrate them in the main stream of the acquisition system in \textit{RFX-mod2}~\cite{garola2018zynqbased, RIGONI2018122}.

\begin{figure}[ht]
\centering
\includegraphics[width=0.48\textwidth]{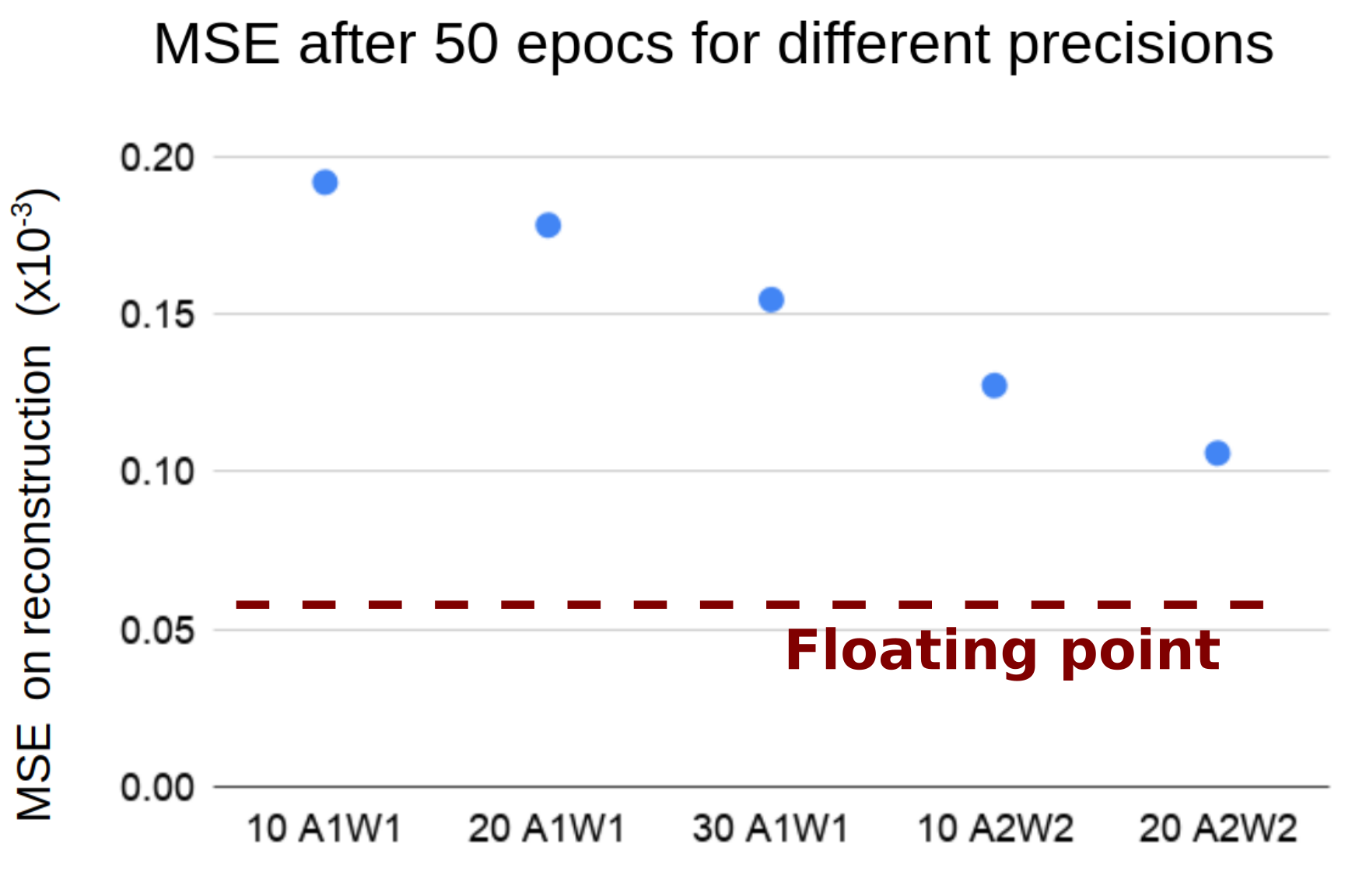}
\caption{Mean squared error (MSE) of the reconstruction computed in a validation set for different topologies and precision in quantized autoencoders. Five configurations have been plotted increasing: the scale factor on the number of neurons per layer (first number on the labels), the activation function number of bits (the digit after A label), and the weight precision bits (the digit after the W label). On the same plot the boundary imposed by a full resolution of the same autoencoder built in floating point.}
\label{fig:7 quantized}
\end{figure}

However these low-end devices come with a poor support for digital signal processing (DSP) computation, making the deployment of these networks unfeasible, either in terms of required DSP units or in time if the same DSP is reused. Even if simple operations are requested, a large number of them are required.
To overcome this limitation we decided to create a low precision version of the encoders, i.e. quantized networks with few bits representing weights and activation functions. The solution proposed here comes from the retraining in a quantization aware framework called \textit{Brevitas}~\cite{alessandro_pappalardo_2019_3525102}, a specific library built on top of \textit{Torch} that implements binarized layers. Obviously the error in the encoder reconstruction is affected by the lower precision of the components that is partially mitigated by an increased amount of nodes per layer. A comparison with different topologies is presented in \textit{Figure}~\ref{fig:7 quantized} that report the corresponding MSE computed on several validation batches of samples. In this case we tested the encoder part of the network, that would be attached at the \textit{DSX3} measure to produce the transmitted codes, to see if such a discretized network is able to reproduce the same results that are based on the standard floating point approach, and if it fits into a simple FPGA component. A template of the network geometry has been set to have an initial 8~bit precision input layer and a set of $S\times[32,32,16,16]$ neurons of low precision layers, where $S$ is a scale factor. The plot labels are composed as ``\textbf{\textit{S}}~W\textbf{\textit{x}}A\textbf{\textit{x}}'', where \textit{S} is the scale used for each training attempt and \textit{x} the precision: ``W1A1'' for the full binary weight and activation, or ``W2A2'' for the 2-bits weight and activation.
~
From the plot it is quite interesting to see how the scale and the precision play on error: for example a double-precision lays on the same trend of a 4x scale of the layer size.

After the training has been completed, and a reasonable precision has been tuned, the second step is to pass it to the logic synthesis.
All the plotted solutions have been visually tested sampling from the validation dataset and seems to reproduce well the input profile, although a more detailed discussion on the reconstruction error should be assessed separating the missing point reconstruction error from the signal noise. Fitting into an average case, the 30 A1W1 case has been chosen to be implemented in FPGA using a recent project from Xilinx called FINN~\cite{Umuroglu_2017}. The encoder part of the \textit{Brevitas} trained network has been first exported in the Open Neural Network Exchange (ONNX) standard~\cite{bai2019} and then fed into FINN creating a streaming pipeline that implements the network in hardware~\cite{umuroglu2018streamlined}.
The actual floating point precision has not been obtained in our simple tests but a very close approximation has been reached. The deployment has been tested on a Xilinx Pynq-Z2 board with a test driver that passes data back and forth via DMA transfer to the logic implementation. Using a 100MHz clock, and a folding factor of 8 for both input and output parallelization, on the FPGA we measured an average latency of $125~\mu s$ (without pipelining). Conversely, on a pipeline with 64 frames FIFO caches (where the application permits that), the observed throughput has been of about $1.3\times10^4$ fps.
This is not yet the highest performance achievable for the binarized layers though, but it still represent a value that is compatible with the time resolution of the observed dynamic that is of about $1~\text{ms}$.
%
%
\section{Conclusions}\label{section:conclusions}

The possible use of DNN for diagnostic data integration in the complex scenario of the active control for fusion relevant plasmas has been studied.
In this context this work proposed a novel set of generative unsupervised models.
In particular a new stochastic approach that provides a robust non-linear embedded representation of the acquired data called Variational Autoencoders.
A study of a lower precision instance of these networks promises a new way to deploy such models in FPGA, opening in turn a chance to apply it to a real-time system.

A general study of these models has been proposed and it has been shown that the learned latent variables of VAE generated embedded manifolds can be considered a practical mean of representing data in a compact yet accurate way. 

A test on a real case scenario based on the reconstruction of the temperature profiles for the \textit{DSX3} diagnostic of \textit{RFX-mod} has been performed.
An accurate model of a $\text{VAE}_6$ has been used as the ground truth to train a further inference network that succeeded to map electromagnetic measures into the temperature profile of the plasma with a good precision. These measure were further added to the composite encoded state of the system and showed to add useful information to the reconstruction of the missing temperature points.

On the other hand the possibility to a concrete hardware implementation was discussed in relation to a new device that is on active development for the electromagnetic signals acquisition upgrade in RFX-mod2.
The hardware deployment of neural network exploits a new quantized representation of the parameters and activation functions of the units. The accuracy issues posed by the reduced precision for them has been tested on a Zynq 7K device. 


\section*{Acknowledgment}
The authors wish to thank Yaman Umuroglu and the \textit{Xilinx FINN} team for the support on network quantization, Paolo Franz and David Terranova for providing all the experimental QSH dataset and the magnetic field reconstructions, Nicola Pomaro for the internal review, and all the staff of the RFX consortium. The FPGA development has been partially supported by the \textit{Xilinx} academic program 2018 in the person of Parimal Patel.

\bibliographystyle{IEEEtran}
\bibliography{main}
\end{document}